\documentclass{ws-ijmpb}

\begin{document}

\markboth{S.A. Gurvitz} {Quantum limit of measurement}

\catchline{}{}{}{}{}

\title{Quantum limit of measurement and the projection postulate}

\author{S.A. Gurvitz}
\address{Department of Particle Physics, Weizmann Institute of Science,
Rehovot 76100, Israel shmuel.gurvitz@weizmann.ac.il}


\maketitle


\begin{abstract}
We study an electrostatic qubit monitored by a point-contact
detector. Projecting an entire qubit-detector wave function on the
detector eigenstates we determine the precision limit for the qubit
measurements, allowed by quantum mechanics. We found that this
quantity is determined by qubit dynamics as well as decoherence,
generated by the measurement. Our results show how the quantum
precision limit can be improved by a proper design of a measurement
procedure.
\end{abstract}



Rapid experimental progress in monitoring of single quantum systems
renewed the interest in the quantum mechanical limitation of
measurement accuracy\cite{rus}. This limitation is originated from
the uncertainty relation between observables of a measured
microscopic system, the so-called standard quantum limit of
measurement\cite{brag}. Yet, a single quantum system is not observed
directly, but through the interaction with a measurement device
(detector). This implies that the quantum limit of measurement is
not related to a measured system only, but rather to an entire
system, including detector\cite{brag}.

A measurement in the quantum mechanical formalism corresponds to
projection of the wave function of an entire system on eigenstates
of the {\em detector}, accessible by an observer. This is the
so-called projection postulate\cite{neu}, analogues to the Bayes
principle in any probabilistic description. Since the detector and
the measured system are interacting, the above projection on the
{\em detector} states affects the measured microscopic system. If
the latter is projected on one of its own states, then the
microscopic system can be measured with any accuracy. However, if
this system is projected on the superposition of its states, the
measurement cannot be precise. Its accuracy is given by a size of
the corresponding wave packet.

In this Letter we apply the above described procedure for a
determination of the precision limit of quantum measurements. As an
example we take an electron trapped inside a double-dot
(electrostatic qubit) and continuous monitored by a point-contact
detector\cite{qb1}. The total system can be treated entirely quantum
mechanically, although the detector represents a macroscopic
(mesoscopic) device\cite{gur}. This allows us to obtain the detector
eigenstates and then to perform projections of the total wave
function on these eigenstates without any additional assumptions.

Consider a qubit, represented by an electron in the double-dot,
placed near a point-contact that separated two reservoirs (the
point-contact detector), Fig.~1. When the first dot, which is far
away from to the point-contact, is occupies (Fig.~1a), the current
through the point-contact is $I_1$. If the second dot, close to the
point contact, is occupied (Fig.~1b), the current decreases
($I_2<I_1$) due to the electrostatic repulsion. The entire system
can be described by the following tunneling Hamiltonian\cite{gur}:
$H=H_{PC}+H_{DD}+H_{int}$, where
\begin{eqnarray}
\label{a1} {\cal H}_{PC}&=&\sum_l E_la_l^\dagger a_l+\sum_r
E_ra_r^\dagger a_r+\sum_{l,r}\Omega_{lr}(a_l^\dagger a_r+H.c.),
\label{a1a}\\
{\cal H}_{DD}&=&E_1 c_1^{\dagger}c_{1}+E_2 c_2^{\dagger}c_{2}+
              \Omega (c_2^{\dagger}c_{1}+ c_1^{\dagger}c_{2})\, ,
\label{a1b}\\
{\cal H}_{int}&=&\sum_{l,r}\delta\Omega_{lr}c_2^{\dagger}
c_2(a^{\dagger}_la_r+H.c.)\, , \label{a1c}
\end{eqnarray}
and $\delta\Omega_{lr}=\Omega'_{lr}-\Omega_{lr}$. Here
$a_{l,r}^\dagger (a_{l,r})$ is the creation (annihilation) operator
for an electron at the level $l$ or $r$ in the left or right
reservoir, and $c^\dagger_{1,2}(c_{1,2})$ is the same operator for
the electron inside the double-dot. $\Omega_{lr}$ is the hopping
amplitude between the states $l$ and $r$ of the reservoirs, and
$\Omega$ is the hopping amplitude between the states $E_1$ and $E_2$
of the qubit. For simplicity we consider electrons as spin-less
fermions. The interaction term $H_{int}$ generates variation of the
hopping amplitude, $\delta\Omega_{lr}={\Omega'}_{lr}-\Omega_{lr}$,
resulting in a decrease of the detector current from $I_1$ to $I_2$,
Fig.~1.
\begin{figure}
{\centering{\psfig{figure=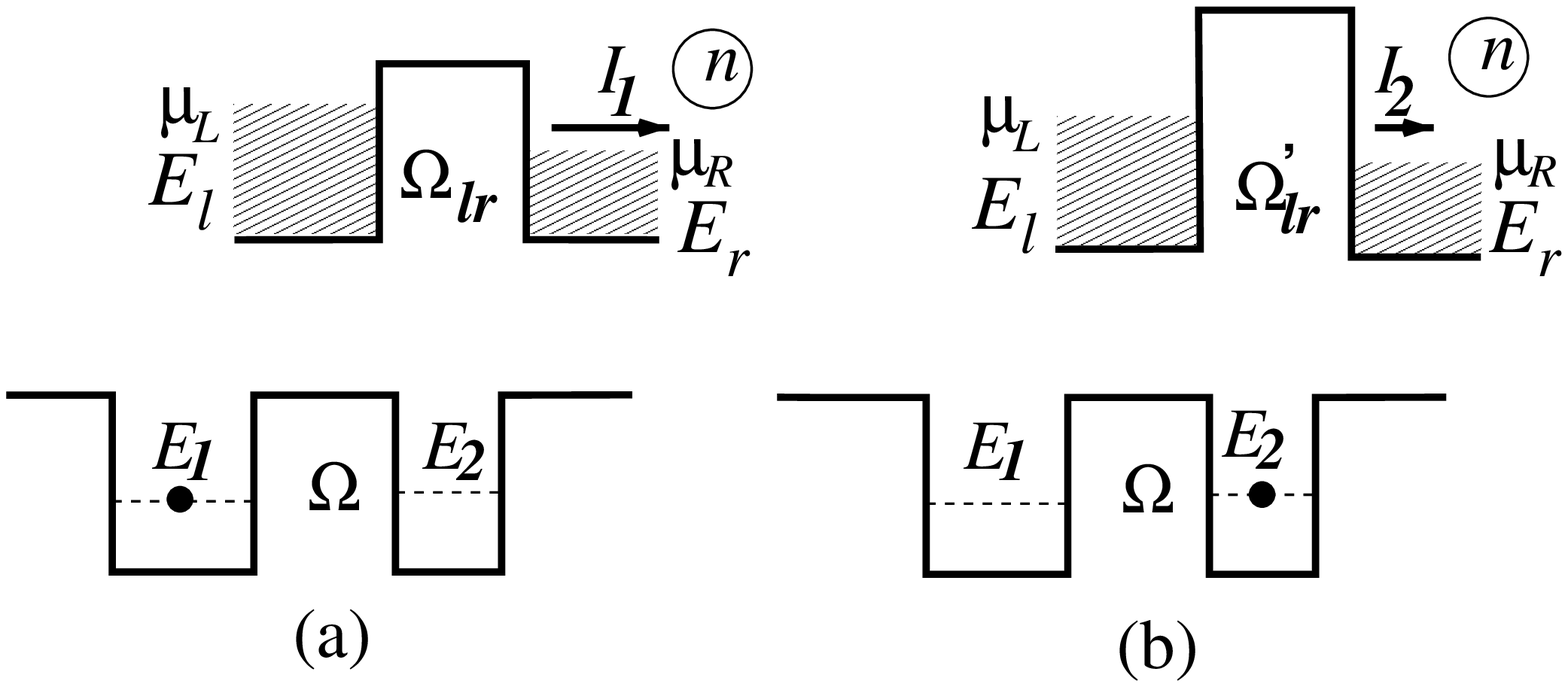,height=5cm,width=12cm,angle=0}}}\\
{\bf Fig.~1:} The electrostatic qubit monitored by the point-contact
detector. The detector current decreases when the left dot of the
qubit is occupied. Here $n$ denotes the number of electrons which
have arrived at the right reservoir by time $t$.
\end{figure}

The wave function describing the entire system can be written as
\begin{eqnarray}
&&|\Psi (t)\rangle = \left [ b_1(t)c_1^{\dagger} + \sum_{l,r}
b_{1lr}(t)c_1^{\dagger}a_r^{\dagger}a_l +\sum_{l<l'\atop r<r'}
b_{1ll'rr'}(t)
c_1^{\dagger}a_r^{\dagger}a_{r'}^{\dagger}a_la_{l'}\right.\nonumber\\
&&\left.+b_2(t)c_2^{\dagger} +\sum_{l,r}
b_{2lr}(t)c_2^{\dagger}a_r^{\dagger}a_l+\sum_{l<l'\atop r<r'}
b_{2ll'rr'}(t)
c_2^{\dagger}a_r^{\dagger}a_{r'}^{\dagger}a_la_{l'}+\cdots \right
]|0\rangle, \label{a2}
\end{eqnarray}
where $b_\alpha (t)$ is the amplitude of finding the system in the
state $\alpha$ determined by the corresponding creation and
annihilation operators. These operators act on the initial
(``vacuum'') state, $|0\rangle$. For simplicity we assume that the
reservoirs are initially at zero temperature and filled with
electrons up to the Fermi energies $\mu_{L,R}$, respectively. All
the amplitudes $b_\alpha (t)$ can be obtained from the
time-dependent Schr\"odinger equation, $\partial_t|\Psi (t)\rangle
=H|\Psi (t)\rangle $.

Now we project the total wave function (\ref{a2}) on the detector
eigenstates. There exists an uncertainty however, about a choice of
the eigenstates, since different detector variables can be
recorded\cite{zur}. In fact, the point-contact detector is not
recorded directly, but via another readout device (``pointer''). The
latter can single out a particular set of the eigenstates through a
coupling with the corresponding detector variables. Yet, in this
work we do not extend our system by including such a pointer in the
Schr\"odinger equation. We assume instead that the detector states
are directly accessible to an ``observer''. Let us examine different
alternatives of the measurement.

Consider first the measurement of number of electrons ($n$) in the
right reservoir (the accumulated charge), Fig.~1. This implies that
the wave function, $|\Psi (t)\rangle$, Eq.~(\ref{a2}), is projected
on the eigenstates of the operator $\hat
N=\sum_ra_{r}^{\dagger}a_{r}$, which can be written as
\begin{equation}
|n\rangle ={\underbrace{a_{r_1}^\dagger a_{r_2}^\dagger\cdots
a_{r_n}^\dagger}_n} |0\rangle\, . \label{a3}
\end{equation}
(Note, that the state $|n\rangle$ is strongly degenerate:
$|n\rangle\equiv |n,\alpha \rangle$, where $\alpha$ corresponds to a
particular configuration of $n$ electrons in the collector). Thus,
$|\Psi (t)\rangle\to {\cal N}^{-1/2} \hat P_n|\Psi (t)\rangle$,
where $\hat P_n=\sum_\alpha |n,\alpha \rangle \langle n,\alpha|$ is
a projection operator and ${\cal N}$ is a normalization factor. One
finds from Eqs.~(\ref{a2}) and (\ref{a3}) that
\begin{eqnarray}
&&\hat P_n|\Psi (t)\rangle = \sum_{l,\ldots\atop r,\ldots }\left
[b_{1{\underbrace{l\ldots}_n}{\underbrace{r\ldots}_n}}(t)
c_1^\dagger a_{\underbrace{r\ldots}_n}^\dagger
a_{\underbrace{l\ldots}_n}+b_{2{\underbrace{l\ldots}_n}{\underbrace{r\ldots}_n}}(t)
c_2^\dagger a_{\underbrace{r\ldots}_n}^\dagger
a_{\underbrace{l\ldots}_n} \right ]|0\rangle \ \ \ \label{a4}
\end{eqnarray}
where the corresponding normalization factor ${\cal
N}=\sum_{l,\ldots ,r,\ldots }[|b_{1l\ldots r}(t)|^2 +|b_{2l\ldots
r}(t)|^2]$ is a probability of finding $n$ electrons in the
collector by time $t$. Eq.~(\ref{a4}) shows that the measurement
leaves the qubit in a linear superposition of its two states.
Therefore one cannot determine the state of the qubit by measuring
the number of electrons in the right reservoirs.

Consider now the measurement of electric current in the right
reservoir. The latter is given by a commutator of $\hat N$ with the
total Hamiltonian of the system, $\hat I=i[H,\hat N]$ (we choose
$e=1$). Using Eqs.~(\ref{a1a})-(\ref{a1c}) we obtain
\begin{equation}
\hat I=i\sum_{l,r}(\Omega_{lr}+\delta\Omega_{lr}c_2^\dagger c_2)
(a_l^\dagger a_r-a_r^\dagger a_l)\equiv \sum_{l,r}\hat I_{lr}
\label{a5}
\end{equation}
The eigenstates of the energy resolved-current operator, $\hat
I_{lr}|I^\pm_{lr}(q)\rangle =I^\pm_{lr}(q)|I^\pm_{lr}(q)\rangle$ are
\begin{equation}
|I^\pm_{lr}(q)\rangle ={1\over\sqrt{2}}(a^\dagger_ra_l\pm
i)c_q^\dagger |0\rangle\, , \label{a6}
\end{equation}
where $q=1,2$ denotes the qubit state, $c^\dagger_q |0\rangle$ and
$I^\pm_{lr}(q)= \pm [\Omega_{lr}+(q-1)\delta\Omega_{lr}]$.
Respectively, the eigenstates of the total current, $|I\rangle$, are
given by a product of $|I^\pm_{lr}\rangle$.

It follows from Eqs.~(\ref{a5}), (\ref{a6}) that by measuring the
energy-resolved current $\hat I_{lr}$ (or the total current, $\hat
I$), one projects the wave function (\ref{a2}) on a certain state of
the qubit, corresponding to an observed value of the current. This
implies that the qubit position can be determined with any accuracy,
in principle, by monitoring directly the detector current (via its
magnetic field\cite{lev}).

If, however, such a direct measurement of the detector current
cannot be performed, one can determine it indirectly, via a
variation of the collector charge. Let us assume that we recorded
$n_0$ electrons in the collector at time $t$. As a result, the
entire system is projected to the state $|n_0\rangle$: $|\psi
(t)\rangle\to \hat P_{n_0}|\psi (t)\rangle$, Eq.~(\ref{a4}), which
is an eigenstate of the operator $\hat N$, Eq.~(\ref{a3}). (We
omitted the index of degeneracy $\alpha$). Next we detect the
accumulated charge $n$ at the time $t+\Delta t$. The final state of
the detector (up to the normalization factor) is
\begin{equation}
\hat P_ne^{-iH\Delta t}\hat P_{n_0}|\Psi (t)\rangle
=|n\rangle\langle\varphi_n(\Delta t)|n_0\rangle\langle n_0|\Psi
(t)\rangle\, , \label{a7}
\end{equation}
where $|\varphi_n(\Delta t)\rangle =\exp (iH\Delta t)|n\rangle$ is
an eigenstate of the operator $\hat N(\Delta t)=\exp (iH\Delta
t)\hat N\exp (-iH\Delta t)$, corresponding to an eigenvalue $n$.
This operator can be expanded in powers of $\Delta t$
\begin{equation}
\hat N(\Delta t)=\hat N +\hat I\Delta t +i[H,\hat I]{(\Delta
t)^2\over 2}+\cdots \, , \label{a8}
\end{equation}
where the current $\hat I$ is given by Eq.~(\ref{a5}). Thus the
time-dependent operator $\hat N(\Delta t)$ includes the qubit
position operator, $c_q^\dagger c_q$, in contrast with $\hat N\equiv
N(0)$. As a result the projection on the eigehstates of $\hat
N(\Delta t)$, Eq.~(\ref{a7}), could determine the qubit position.

Let us take small  $\Delta t$ (``measurement time'') in
Eq.~(\ref{a8}) such that $\hat N(\Delta t)\simeq \hat N +\hat
I\Delta t$. If $\hat N$ and $\hat I$ commute, then the eigenstates
of $\hat N(\Delta t)$ would be a product of eigenstates of these
operators: $|\varphi_n(\Delta t)\rangle=|n',I(q)\rangle$, where
$q=\{ 1,2\}$ denotes the qubit state and $n=n'+I(q)\Delta t$. It
follows from Eq.~(\ref{a7}) that $n'=n_0$ and $I(q)\equiv I_{\Delta
n}(q)=\Delta n/\Delta t$, where $\Delta n=n-n_0$. As a result the
qubit is projected in the state $q$ corresponding to the variation
of the collector charge, $\Delta n$.

In fact, the operators $\hat N$ and $\hat I$ do not commute. In this
case it is only the average current, $\overline{I}_{\Delta
n}=\overline{\Delta n}/\Delta t$ which can be determined from
ensemble measurements of $\Delta n$, where its dispersion, $\delta
I_{\Delta n}=[\overline{(\Delta n)^2}-(\overline{\Delta
n})^2]^{1/2}/\Delta t$, diverges as $(\Delta t)^{-1/2}$ for $\Delta
t\to 0$. The latter restricts the accuracy of the qubit
measurements, respectively.

The measurement accuracy, however, increases by increasing the
measurement time, since $\delta I_{\Delta n}(q)\to 0$ for $\Delta
t\to \infty$\cite{fn1}.  Yet, this can be done only if the qubit is
not ``moving'', (i.e. the hopping amplitude $\Omega =0$, Fig.~1).
Then $[H,\hat I]=0$, so that the higher order terms in the expansion
(\ref{a8}) vanish. If it is not the case ($\Omega\not =0$), the
current it driven by the qubit, so that $[H,\hat I]$ and the higher
order commutators in Eq.~(\ref{a8}) are not zero. The average
contribution from these terms to $I_{\Delta n}$, which we denote as
$\delta_1 I_{\Delta n}(\Delta t)$, increase with $\Delta t$. This
suggests that the quantum limit of the qubit measurement is
determined by the optimal measurement time ($\Delta t$) which
minimizes the total error,
\begin{equation}
[\delta_2 I_{\Delta n}(\Delta t)]^2 =[\delta I_{\Delta n}(\Delta
t)]^2+[\delta_1 I_{\Delta n}(\Delta t)]^2\, . \label{aa8}
\end{equation}

In order to perform this procedure we introduce the density matrix
$\sigma_{qq'}^{(n)}(t)=\langle n,q'|\Psi (t)\rangle\langle\Psi
(t)|n,q\rangle$, where the wave function $|\Psi (t)\rangle$ is given
by Eq.~(\ref{a2}). It was demonstrated in\cite{gur,gur1} that for
large large bias voltage, $V=\mu_L-\mu_R$ (Fig.~1), the
time-dependent Schr\"odinger equation, $\partial_t|\Psi (t)\rangle
=H|\Psi (t)\rangle$, can be reduced to the following Bloch-type rate
equations for the density matrix $\sigma_{qq'}^{(n)}(t)$ by assuming
weak energy dependence of the transition amplitudes
($\Omega_{lr}=\bar\Omega,~ \Omega'_{lr}=\bar\Omega'$)\cite{gur},
\begin{eqnarray}\label{a10a}
\dot\sigma_{11}^{(n)} &=&
-D_1\sigma_{11}^{(n)}+D_1\sigma_{11}^{(n-1)} +i\Omega
(\sigma_{12}^{(n)}-\sigma_{21}^{(n)})\\ \label{a10b}
\dot\sigma_{22}^{(n)}
&=&-D_2\sigma_{22}^{(n)}+D_2\sigma_{22}^{(n-1)} -i\Omega
(\sigma_{12}^{(n)}-\sigma_{21}^{(n)})
\\ \label{a10c}
\dot\sigma_{12}^{(n)} &=& i(E_2-E_1)\sigma_{12}^{(n)}+
i\Omega(\sigma_{11}^{(n)}-\sigma_{22}^{(n)})
-\frac{D_1+D_2}{2}\sigma_{12}^{(n)}
+(D_1D_2)^{1/2}\sigma_{12}^{(n-1)}
\end{eqnarray}
where $D_{1,2}=T_{1,2}V$ and $T_{1,2}$ is the transmission
probability of the barrier: $T_1=(2\pi)^2\bar\Omega^2\rho_L\rho_R$
and $T_2=(2\pi)^2(\bar\Omega')^2\rho_L\rho_R$, where
 $\rho_{L,R}$ is the density of states in the left (right)
reservoir, Fig.~1.

Solving Eqs.~(\ref{a10a})-(\ref{a10c}) one can find all quantities
needed for evaluation of $\delta I_{\Delta n}$ and
$\delta_1I_{\Delta n}$. For instance, in order to evaluate the
average value of $\Delta n$ and its dispersion we have to solve
these equations with the initial condition $n=n_0$. It follows from
Eqs.~(\ref{a10a})-(\ref{a10c}) that the density matrix
$\sigma_{qq'}^{(n)}(t)$ depends only on $\Delta n=n-n_0$. Thus we
can take $n_0=0$ and $\Delta n =n$. Then the average values
$\overline{\Delta n}=\overline{n}$ and $\overline{(\Delta
n)^2}=\overline{n^2}$ are given by
\begin{eqnarray}
\overline{n}(t)&=&\sum_nnP_n(t)=\overline{n}_{11}(t)+\overline{n}_{22}(t)\,
, \label{a11a}\\
\overline{n^2}(t)&=&\sum_nn^2P_n(t)=\overline{n_{11}^2}(t)+\overline{n_{22}^2}(t)\,
, \label{a11b}
\end{eqnarray}
where $P_n(t)=\sigma_{11}^{(n)}(t)+\sigma_{22}^{(n)}(t)$ is the
probability of finding $n$ electron in the collector. Multiply
Eqs.~(\ref{a10a})-(\ref{a10c}) by $n$ and sum over $n$ one finds
\begin{eqnarray}
\dot{\overline{n}}_{11} &=& D_1\sigma_{11}
+i\Omega (\overline{n}_{12}-\overline{n}_{21})\\
\label{a12a} \dot{\overline{n}}_{22} &=& D_2\sigma_{22}
-i\Omega (\overline{n}_{12}-\overline{n}_{21})\\
\label{a12b} \dot{\overline{n}}_{12} &=&
i(E_2-E_1)\overline{n}_{12}+ i\Omega(\overline{n}_{11}-
\overline{n}_{22}) -\frac{\Gamma_d}{2}
\overline{n}_{12}+(D_1D_2)^{1/2}\sigma_{12}\, , \label{a12c}
\end{eqnarray}
where $\sigma_{qq'}(t)=\sum_n\sigma_{qq'}^{(n)}(t)$ is the qubit
density matrix, and $\Gamma_d=(\sqrt{D_1}-\sqrt{D_2})^2$ is the
decoherence rate\cite{gur}.

Similarly multiplying Eqs.~(\ref{a10a})-(\ref{a10c}) by $n^2$ and
sum over $n$ one obtains
\begin{eqnarray}\label{a13a}
\dot{\overline{n_{11}^2}} &=&2 D_1\overline{n}_{11}+
 D_1\sigma_{11}
+i\Omega (\overline{n_{12}^2}-\overline{n_{21}^2})\\ \label{a13b}
 \dot{\overline{n_{22}^2}} &=&2 D_2\overline{n}_{22}+
 D_2\sigma_{22}
-i\Omega (\overline{n_{12}^2}-\overline{n_{21}^2})\\ \label{a13c}
 \dot{\overline{n_{12}^2}}&=&
i(E_2-E_1)\overline{n_{12}^2}+ i\Omega(\overline{n_{11}^2}-
\overline{n_{22}^2}) -\frac{\Gamma_d}{2}
\overline{n_{12}^2}+(D_1D_2)^{1/2}(2\overline{n}_{12}+\sigma_{12})\,
.
\end{eqnarray}

The qubit density matrix $\sigma_{qq'}(t)$ can be easily found from
Eqs.~(\ref{a10a})-(\ref{a10c}) by tracing it over $n$,
\begin{eqnarray}\label{aa10a}
\dot\sigma_{11} &=&i\Omega (\sigma_{12}-\sigma_{21})\\ \label{aa10b}
\dot\sigma_{12} &=& i\epsilon_{21}\sigma_{12}+
i\Omega(2\sigma_{11}-1) -\frac{\Gamma_d}{2}\sigma_{12}\, ,
\end{eqnarray}
and $\sigma_{22}(t)=1-\sigma_{11}(t)$. This quantity,
$\sigma_{qq'}(t)$, determines the detector average current,
$\overline{I}(t)=\langle\Psi (t)|\hat I|\Psi (t)\rangle$. Indeed, it
follows from Eqs.~(\ref{a2}), (\ref{a5}) that
\begin{equation}
\overline{I}(t)=D_1\dot{\overline{n}}_{11}(t)+D_2\dot{
\overline{n}}_{22}(t)=D_2+\Delta D\, \sigma_{11}(t)\, , \label{a14}
\end{equation}
where $\Delta D=D_1-D_2$ is an average ``signal''. Obviously,
$\overline{I}(\Delta t)=\overline{n}/\Delta t\equiv
\overline{I}_n(\Delta t)$ for small $\Delta t$. The dispersion of
$I_n(\Delta t )$ can be found from Eqs.~(\ref{a11b}),
(\ref{a13a})-(\ref{a13c}),
\begin{equation}
(\delta I_n)^2=\overline{n^2}(\Delta t)/(\Delta
t)^2-\overline{I}_n^2\simeq I_n(0)/\Delta t\, . \label{a15}
\end{equation}
As expected, $\delta I_n(\Delta t)$ diverges as $1/\sqrt{\Delta t}$
for $\Delta t\to 0$.

Let us evaluate the contribution from higher order terms in the
expansion (\ref{a8}), which generate variation of the average
current, $\delta_1 I_n$. (In our case this variation is produced by
the qubit only). Therefore, one can write $\delta_1I_n(\Delta
t)=|\overline{I}_n(\Delta t)-\overline{I}_n(0)|$. Using
Eq.~(\ref{a14}) we evaluate this quantity as
\begin{equation}
\delta_1 I_n =\Delta D\,|\sigma_{11}(\Delta t)-\sigma_{11}(0)|\, ,
 \label{a16}
\end{equation}
where $\sigma_{11}(t)$ is obtained from
Eqs.~(\ref{aa10a})-(\ref{aa10b}). For instance, for aligned levels,
$E_1=E_2$, and $\sigma_{11}(0)=1$,
\begin{equation}
\delta_1 I_n={\Delta D\over2}\left |e^{-{\Gamma_d\over 4}\Delta
t}\left[\cos (\omega \Delta t)+\eta\sin (\omega \Delta
t)\right]-1\right |\, ,
 \label{a17}
\end{equation}
where $\eta=\Gamma_d/4\omega$ and $\omega
=2\Omega\sqrt{1-(\Gamma_d/8\Omega )^2}$ is the Rabi frequency. As
expected, $\delta_1 I_n\to 0$ for $\Omega\to 0$.

Eqs.~(\ref{a15}) and (\ref{a17}) allow us to evaluate the optimal
$\Delta t$ by minimizing $[\delta_2 I_n(\Delta t)]^2$,
Eq.~(\ref{aa8}). We take for simplicity $\Delta D\ll D$, where
$D=(D_1+D_2)/2$. As a result $\Gamma_d\simeq (\Delta D)^2/4D$. We
first consider weak distortion of the qubit, $\Gamma_d/8\ll\Omega$.
Then expanding Eq.~(\ref{a17}) in powers of $\Delta t$ we easily
obtain for the optimal measurement time and for the corresponding
measurement limit
\begin{equation}
\Delta t= {1\over 2\Omega}\, \left({2\Omega\over\Gamma_d}\right
)^{1/5},~~~(\delta_2 I_n)^2={5D\Omega \over 2}\left({\Gamma_d\over
2\Omega}\right )^{1/5}\, .
 \label{a18}
\end{equation}

In order to observe Rabi oscillations of the qubit in a single
measurement run one needs $\delta_2 I_n\ll \Delta D$, at least. It
follows from Eq.~(\ref{a18}) that this condition corresponds to
$\Omega\ll 2\Gamma_d$. This, however cannot be combined with the
condition of weak qubit distortion, used in Eq.~(\ref{a18}). Hence,
one cannot observe Rabi oscillations in a single run, but only in an
ensemble average of different runs.

Consider now large decoherence limit, $\Gamma_d/8\gg\Omega$. Then
the qubit is strongly affected by the detector. As a result, the
electron stays in the same dot for a long time,
$t\sim\Gamma_d/8\Omega^2$ (``quantum Zeno'' effect). Indeed, one
finds from Eqs.~(\ref{aa10a})-(\ref{aa10b}) that $\sigma_{11}=
[1+\exp(-8\Omega^2t/\Gamma_d)]/2$ for $t\gg 1/\Gamma_d$.
Respectively, the optimal measurement time and the measurement limit
are given by
\begin{equation}
\Delta t= {1\over 4\Omega }\, \left ({\Gamma_d\over 2\Omega}\right
)^{1/3},~~~(\delta_2 I_n)^2=6D\Omega \left ({2\Omega
\over\Gamma_d}\right )^{1/3}\, .
 \label{a19}
\end{equation}
In contrast with the previous case, Eq.~(\ref{a18}), the measurement
time $\Delta t$ increases with $\Gamma_d$. This is not surprising
since large decoherence generated by the detector, localizes the
qubit for a long time. Therefore it behaves as a static one so that
the measurement time increases.

It follows from our arguments that the quantum precision limit
depends on a particular set of the detector observables which the
total wave function is projected on. In this Letter we discussed two
alternative sets related to charge and current states of the
point-contact detector. It was demonstrated that single projection
on charge states cannot measure the qubit state. However, two
consecutive projections of the entire system on the charge states
can measure the qubit state, but only with a limited accuracy. On
the other hand, direct projection on the current state of the
detector can determine the qubit state with absolute precision, in
principle. If this could be realized, one would arrive to the Zeno
paradox\cite{zeno}, i.e. to complete freezing of a system as a
result of continuous measurement.

This shows a necessity of including a ``pointer'' in the total
Hamiltonian coupled with current states of the detector (von Neumann
hierarchy\cite{neu}). In this case two consecutive projections of
the total wave function on the pointer states, used for a
determination of the detector current, would restrict the
measurement accuracy in a total analogy with the previous case. If
the pointer is coupled with charge states of the detector, it
obviously cannot decrease the quantum measurement limit found in our
calculations. We assume also that the pointer cannot essentially
increase this limit, since otherwise the von Neumann hierarchy of
measurements would not converge. This problem, however needs a
special attention.

Our final results on quantum limit of measurement involved only
average quantities, which were obtained without any explicit resort
to the projection postulate. Indeed, the latter is related only to a
single measurement. Nevertheless, as we demonstrated in this Letter,
the use of the projection postulate was very useful in a
determination of quantum limit of measurement. In particular, it
clearly displayed the detector variables  which would allow us to
measure a microscopic system with maximal accuracy. Such variables
are usually represent commutators (time derivatives) of operators
describing the detector states.

An appropriate choice of this variable depends of a particular
measurement apparatus. For instance, for a single electron
transistor (SET) detector, one needs to use the second commutator
(``acceleration'') of the accumulated charge. In contrast with the
point-contact detector, the projection to current states of the SET
would not determine the qubit state. The measurement of the charge
``acceleration'' can be very useful if the corresponding operator
would commute with the charge operator. In this case one can design
an appropriate procedure of projecting on the charge states at
different times which would diminish the quantum measurement limit.
This however must be a subject of a separate investigation.





\begin{thebibliography}{0}
\bibitem{fn1} In this limit $\hat N(\Delta t)=\hat
N+\hat I\Delta t$ is dominated by the second term. As a result, the
measurement of $\Delta n$ determines the current $I$, and therefore
the qubit is projected in one of its states.
\end{thebibliography}
\end{document}